\documentclass{article}

\usepackage{arxiv}

\usepackage[utf8]{inputenc} 
\usepackage[T1]{fontenc}    
\usepackage{hyperref}       
\usepackage{url}            
\usepackage{booktabs}       
\usepackage{amsfonts}       
\usepackage{nicefrac}       
\usepackage{microtype}      
\usepackage{lipsum}
\usepackage{graphicx}
\usepackage{color,soul}
 \usepackage{placeins}
\usepackage{caption}
\usepackage{subcaption}

\graphicspath{ {./images/} }

\title{Piezoelectric modulus prediction using machine learning and graph neural networks}

\author{
Jeffrey Hu \\
  Dutch Fork High School\\
 1400 Old Tamah Rd, Irmo, SC 29063 \\
  \And
 Yuqi Song\\
 Department of Computer Science and Engineering\\
  University of South Carolina\\
  Columbia, SC 29201 \\
  \texttt{yuqis@email.sc.edu} \\
}

\begin{document}
\maketitle
\begin{abstract}
Piezoelectric materials are widely used in all kinds of industries such as electric cigarette lighters, diesel engines and x-ray shutters. However, discovering high-performance and environmentally friendly (e.g. lead-free) piezoelectric materials is a difficult problem due to the sophisticated relationships from materials' composition/structures to the piezoelectric effect. Compared to other material properties such as formation energy, band gap, and bulk modulus, it is much more challenging to predict piezoelectric coefficients. Here, we propose a comprehensive study on designing and evaluating advanced machine learning models for predicting the piezoelectric modulus from materials' composition and/or structures. We train the predicted model based on extensive feature engineering combined with machine learning models (Random Forest and Support Vector Machines) and automated feature learning based on deep graph neural networks. Our SVM model with crystal structures feature outperform other methods, we also use this model to predict the piezoelectric coefficients for 12,680 materials from the Materials Project database and report the top 20 potential piezoelectric materials.
\end{abstract}


\section{Introduction}



\begin{figure}[ht!]
  \centering
      \includegraphics[width=0.75\textwidth]{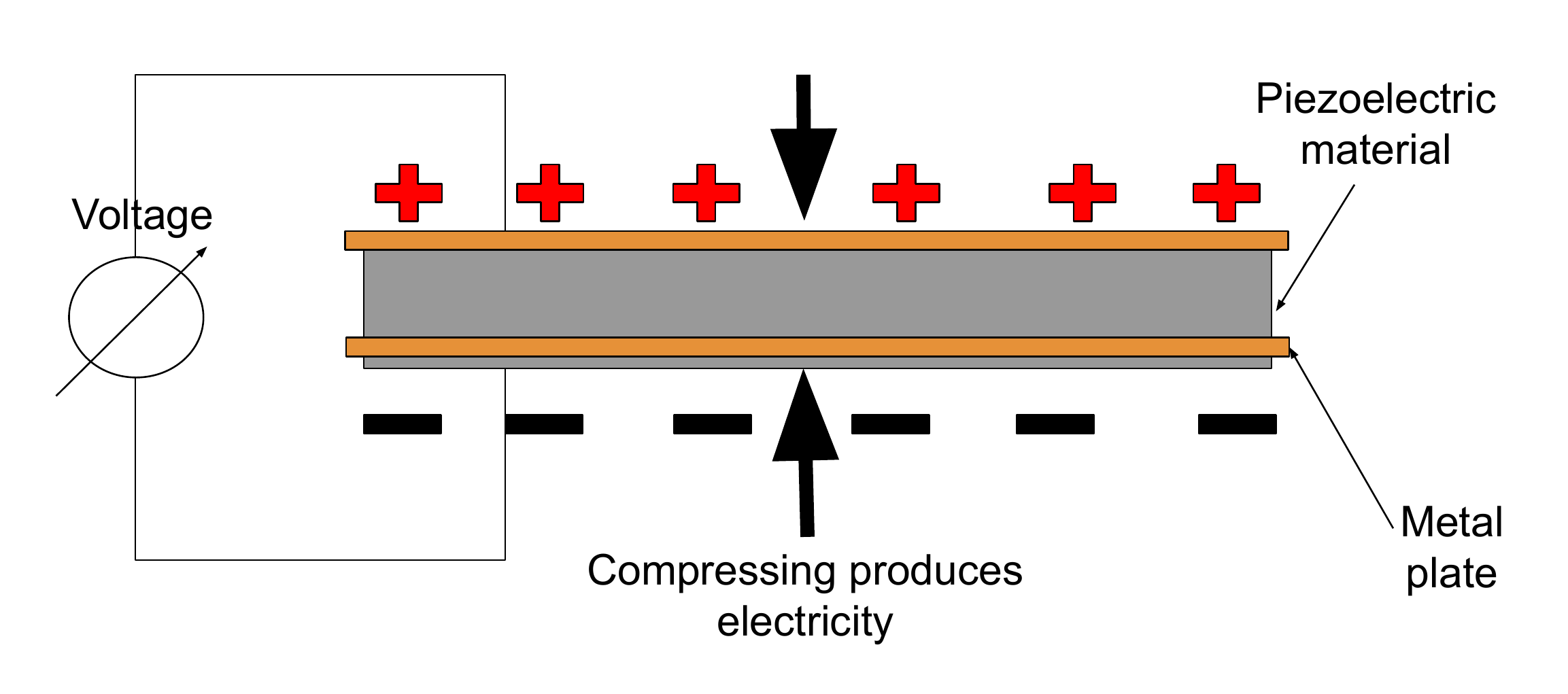}
  \caption{Piezoelectric effect. }

  \label{fig:effect}
\end{figure}

\begin{figure}[ht!]
  \centering
      \includegraphics[width=0.75\textwidth]{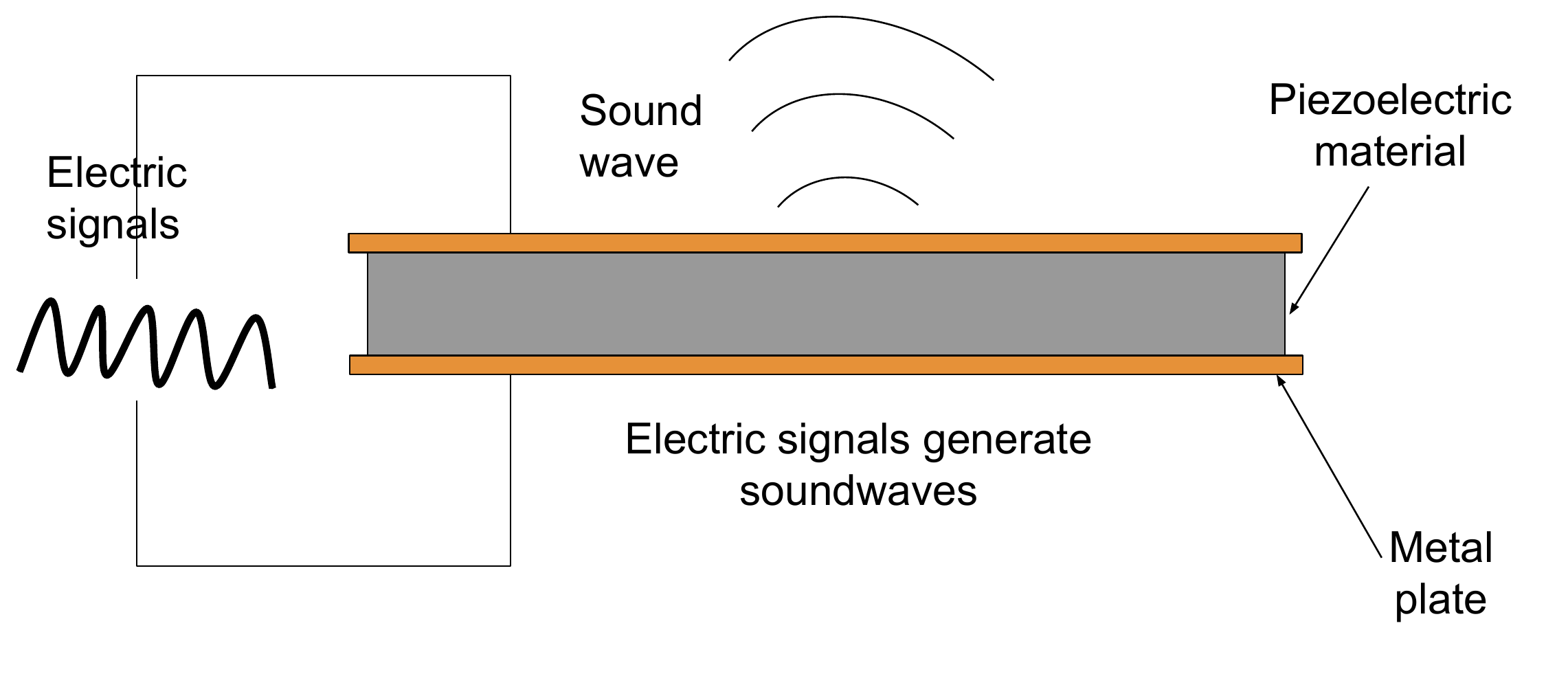}
  \caption{Inverse piezoelectric effect.}

  \label{fig:inverseppp}
\end{figure}

Piezoelectric materials are materials that can generate charge from applied stress (Figure \ref{fig:effect}). These materials are also able to exhibit the inverse piezoelectric effect (as shown in Figure \ref{fig:inverseppp}) which is the generation of mechanical strain reacting to an applied electrical field \cite{nanthakumar2016detection,do2019isogeometric}. These two unique properties have enabled a variety of applications such as ultrasonic detectors, microphones, sonar devices, and ignition systems, which are all based on the piezoelectric effect \cite{piezoinfo}. The piezoelectric materials themselves are used in many daily appliances such as electric cigarette lighters, gas grills and burners, and cold cathode fluorescent lamps, electric guitars, electronic drum pads, and medical acceleromyography. Piezoelectric materials can also used as actuators for accurately positioning objects, which is useful in loudspeakers, piezoelectric motors, laser electronics, inkjet printers, diesel engines, and x-ray shutters. 

In general, there are three types of piezoelectric materials: naturally occurring, man-made, and ceramic materials. Naturally occurring crystals include quartz, sucrose, Rochelle salt, topaz, tourmaline, and berlinite (AlPO$_4$). Man-made piezoelectric crystals include gallium orthophosphate (GaPO$_4$) and langasite (LA$_3$Ga$_5$SiO$_14$). Ceramic piezoelectric materials include BaTiO$_3$, PbTiO$_3$, PZT, KNbO$_3$, LiNBO$_3$, LiTaO$_3$, Na$_2$WO$_4$. PZT is currently the most commonly used piezoelectric ceramic, however, it is not environmentally friendly so alternative materials were developed in response which include sodium potassium niobate (NaKNb), bismuth ferrite (BiFeO\textsubscript{3}), sodium niobate (NaNbO\textsubscript{3}). New applications such as energy harvesting call for development of new types of lead free piezoelectric materials. Despite the importance of piezoelectric materials, there is no solid understanding of how the crystal structure determines the piezoelectric modulus and how to design materials with higher piezoelectric coefficients.

In this work, we aim to develop a machine learning (ML) model for accurate prediction of piezoelectric modulus so that it can be used for screening novel environmentally friendly piezoelectric materials. The piezoelectric coefficient or piezoelectric modulus, usually written d33, measures the volume change when a piezoelectric material is subject to an electric field or the polarization on application of a stress. There have been a variety of machine learning models developed for materials property predictions such as formation energy, band gaps \cite{na2020tuplewise,gladkikh2020machine}, fermi energy \cite{xie2018crystal}, hardness \cite{mazhnik2020application}, Poisson's ratios, elastic (shear/bulk moduli) \cite{xie2018crystal,zhao2020predicting,revi2021machine}, superconductor transition temperature \cite{meredig2018can,stanev2018machine,li2020critical,dan2020computational,revathy2021prediction}, ion conductivity \cite{dunn2020benchmarking,guan2020resolving,sendek2018machine,hatakeyama2019synthesis}, flexoelectricity \cite{ghasemi2017level, ghasemi2018multi,do2019isogeometric} and etc. These ML models can be categorized into three main categories in terms of the input information: composition descriptors based models, structure information based models, and hybrids. See \cite{dunn2020benchmarking} for a comprehensive set of descriptors and related ML models. With sufficient amount of dataset, it has been shown that composition based ML models alone can achieve highly accurate models for formation energy \cite{ward2016general,jha2018elemnet} and band gap predictions \cite{zhuo2018predicting}. Actually, some of those high-performance reports of composition based ML models are very likely  due to the high redundancy of the test sets as regard to the training set when random splitting or cross-validation evaluations are used for large datasets such as Materials Project. These datasets tend to have many highly similar samples due to the tinkering discovery and study process of materials over history. Three recent solid benchmark evaluations have clearly shown that most of time the structure based prediction models outperform those composition models \cite{dunn2020benchmarking,bartel2020critical,fung2021benchmarking}. For example, Bartel et al. \cite{bartel2020critical} showed that composition models failed to properly distinguish relative stability of inorganic materials. Instead they found that including structure in the representation can lead to non-incremental improvement in stability predictions (with the use of CGCNN graph neural network), which serves as a strong endorsement for structural models. Even with this range of ML models for diverse materials properties, there is currently no study on ML prediction of piezoelectric coefficients in the literature.

Appropriately representing materials' characteristics is an important and necessary task in machine learning method. In general, material features \cite{schleder2019dft} can be divided into composition features, structure features, and other more complex representation calculated by specific models. Composition features mainly include chemical element stoichiometric information such as atomic element types, the number of elements, atomic weight; Magpie \cite{ward2016general} is a common used composition feature set. Structural features focus on crystal system information and present the chemical species and atomic coordinates instance. Typical structural features include Coulomb matrix, Ewald sum matrix, sine matrix, Many-body Tensor Representation (MBTR), Atom-centered Symmetry Function (ACSF), and Smooth Overlap of Atomic Positions (SOAP) as implemented in the Dscribe library \cite{himanen2020dscribe}. There are more complex representation from molecule-oriented features to descriptors for extended materials systems and tensorial properties, such as symmetrized gradient-domain machine learning (sGDML) \cite{chmiela2018towards}. However, a recent benchmark study shows that representation learning enabled by graph neural networks such as crystal graph convolutional neural networks (CGCNN) \cite{xie2018crystal} tend to greatly outperform traditional heuristic features such as SOAP features. Other deep representation learning  models have also been applied to grid or voxel like representations for materials property prediction and generations \cite{kajita2017universal,samaniego2020energy,zhao2020predicting}.


Currently, there are 1,705 materials in the Materials Project database with measured piezoelectric modulus. In this paper, based on those known piezoelectric materials ,we aim to train a machine learning model which can predict piezoelectric coefficients with good performance. Here, we explore 5 types of features to train random forest models and support vector machines models. Besides, 5 graph neural networks combined with composition and structural features are also trained. Their performances are evaluated by k-fold cross validation in experiments. Finally, we apply the trained SVM model to predict 12,680 materials' piezoelectric coefficients and report top 20 potential piezoelectric materials.

\section{Methods}
\label{sec:headings}

\subsection{Dataset}

We download the piezoelectric coefficient dataset from the Materials Project (MP) database which contained 1,705 inorganic materials. In the dataset, 65 samples have 0 as the piezoelectric coefficient (PC). There are only 8 materials with a coefficient greater than 20 C/$m^2$ including: AgBiO$_3$(24.84), Sm$_2$CdSe$_4$(32.504), Li$_4$CO$_4$(33.35), MnCO$_3$(39.83), Na$_4$CO$_4$(50.45), Ba(Si$_3$N$_4$)2(67.67), CdGeO$_3$(75.01), Pr$_3$NF$_6$(86.09).

\begin{figure}[ht!]
  \centering
      \includegraphics[width=0.75\textwidth]{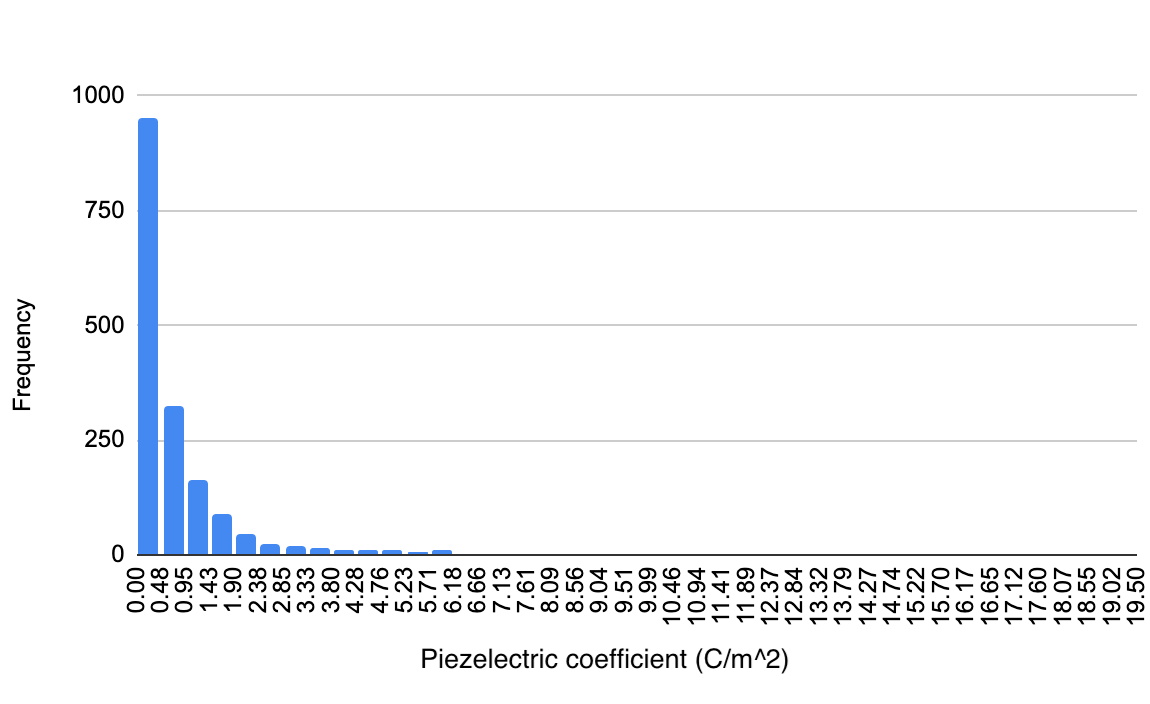}
  \caption{Piezelectric coefficient distribution of the MP dataset. For better visualization, we have excluded the 8 samples with piezoelectric coefficent >20 $C/m^2$.}
  \label{fig:histogram}
\end{figure}

Based on the graph shown in Figure \ref{fig:histogram}, the piezoelectric coefficient values are highly unbalanced with very few samples with high piezoelectric coefficients which makes their prediction to be very challenging.

\subsection{Features}
Selecting a set of appropriate material features is critical to train a successful machine learning model for predicting piezoelectric properties, which are very unique and difficult to predict as we found during our research process. To solve this problem, we try a wide range of different features including the following: 

\begin{itemize}
    \item Composition based magpie features \cite{ward2016general}: these are elemental based property features which are composed of 6 statistics (mean, mean absolute deviation, range, minimum, maximum, and mode) of a set of 22 elemental properties including atom number, Mendeleev number, atomic weight, melting temperature, periodic table column, periodic table row, covalent radius, electronegativity, Ns valence(number of s orbital valence), Np valence, Nd valence, Nf valence, N valence, Ns unfilled, Np unfilled, Nd unfilled, Nf unfilled, N unfilled, GS volume pa, GS band gap, GS magnetism moments, space group number.
    \item Oxidation states features: these are statistics (maximum, range, standard deviation) about the oxidation states for each species.
    \item Structural features: we found that the symmetry degree of the crystal structure has strong effects on the piezoelectric effects so we defined a set of crystal structural features based on the crystal systems of the materials. These features include: (1) number of perpendicular face pairs of the unit cell, (2) the number of equal edges of the unit cell, (3) one hot encoding of crystal systems (cubic, hexagonal, monoclinic, orthorhombic, tetragonal, triclinic, trigonal). We also included the following features: (1) maximum atom radius, (2) minimum atom radius, (3) average atom radius, (4) geometric mean of atom radius, (5) standard deviation of atom radius, (6) maximum Poisson's ratio, (7) minimum Poisson ratio, (8) average Poisson's ratio, (9) geometric mean of Poisson's ratio, (10) standard deviation of Poisson's ratio.
    \item Energy and magnetism features: (1) e\_above\_hull, (2) band gap, (3) total magnetization of magnetism, and (4) total magnetization. These features were downloaded from the Materials Project database.
    \item Elastic modulus features: (1) shear modulus which indicates an object's tendency to shear when acted upon by opposing forces, (2) bulk modulus, it describes volumetric elasticity.
    \item Raw crystal structures for graph neural network based feature learning.
\end{itemize}


\subsection{Machine learning algorithms}
\paragraph{Random Forest}
Random forest (also known as random decision forest) is a type of learning method specifically made for classification and regression which runs on the construction of decision trees during the training process. Advantages of random forest include its ability to reduce over fitting in decision trees which improves accuracy and its compatibility with different types of values. Another useful feature of the random forest model is that it can rank the importance of variables in regression tasks effectively.In the training part, we set the hyper-parameter number of tree as 200 and the number of estimators as 50.

\paragraph{Support Vector Machines}
The support vector machine algorithm is another machine learning algorithm for classification, regression, and other tasks that works by creating a hyper plane(s) in a high- or infinite-dimensional space. The advantages of support vector machines are that they are very effective in high dimensional spaces, especially when the number of dimensions is greater than the number of samples. This method has been widely used in material science research \cite{tang2010prediction, abd2017modelling, liu2020modelling}.In our model, we select rbf kernel, and we set the regularization parameter c = 1.0 and epsilon = 0.2.

\paragraph{Graph Neural Networks}
Graph neural network \cite{zhou2020graph} is a machine learning model that directly takes graphs (composed of vertexes and edges) as an input. Graph neural networks have wide applications in various domains such as social networks, knowledge graphs, recommender systems, and life science. One of the major advantages of graph neural networks is its capability to learn or model dependencies (interactions) between nodes in a graph which is highly suitable for modeling interactions between atoms in materials \cite{xie2018crystal}.

In this paper, we use five graph neural networks including: SchNet \cite{schutt2017schnet}, CGCNN \cite{xie2018crystal}, MPNN \cite{gilmer2017neural}, MEGNET \cite{chen2019graph}, and graph attention neural network(GATGNN) \cite{louis2020graph} for piezoelectric coefficient predictions. These models have recently been evaluated in a benchmark study with comparable performances \cite{fung2021benchmarking}. In our experiments, we use those models' default parameters. 

\paragraph{SchNet} SchNet is a type of deep neural network used for predicting molecular energies and atomic properties. This network observes physical laws and achieves rotation and translation invariance. Interactions between atoms are described by three interaction blocks. This model adopts continuous-filter convolutional layers as the main building block for neural network architecture which allows it to model local correlations without requiring the data to lie on the grid.

\paragraph{CGCNN} CGCNN (Crystal Graph Convolutional Neural Network) is a type of deep graph neural network used to learn material properties from the relationships of atoms within a crystal which gives in depth representations of crystalline materials. CGCNN is flexible in predicting properties and accurately extracting information of different materials. It has been successfully applied to predict a variety of materials properties such as formation energy, absolute energy, Fermi energy, band gaps, bulk/shear moduli, and Poisson's ratio. Among them, the bulk/shear moduli and Posisson ratio prediction models are trained with only 2041 samples, which is close to the dataset size in our study.

\paragraph{MPNN} MPNN (Message Passing Neural Networks) is a type of neural network that accurately predicts important molecular properties of materials. MPNN is favorable in predicting molecular properties with relatively high accuracy. This model is invariant to graph isomorphism, which is composed of message functions, vertex update functions, and readout functions which are all differential functions. MPNN generally works on directed graphs with separate channels for incoming and outgoing edges. 

\paragraph{MEGNET} MEGNET (MatErials Graph Network) is a new type of graph neural network algorithm for material property prediction that uses two new strategies to address the data scarcity problem. One strategy is to build single free energy model by incorporating the temperature, press, and entropy as global state inputs. MEGNet models are found to outperform previous ML models in predicting properties and achieve higher accuracy dealing with larger datasets. MEGNET is high performing in targeting a wide variety of properties for both molecules and crystals. MEGNet models use learned element embedding that encode periodic chemical trends which can be learned from other property models trained on larger datasets.

\paragraph{GATGNN} This graph neural network model is composed of multiple graph-attention layers (GAT) and global attention layers. These GAT layers enable this model to efficiently learn complex bonds shared among atoms in each atom's local environments. Using this model, we aim to train machine learning models to learn the structural features for piezoelectric coefficient predictions. 

The graph neural network models used here are adapted from the benchmark study in \cite{fung2021benchmarking} except that we have added the differentiable group normalization operator \cite{zhou2020towards} into above five neural network models along with skip-connection which is first proposed in ResNet and recently in graph neural networks. Such modifications have allowed us to train graph neural network models to achieve better performance. We use the default parameters for training these models with epoch number of 250.

\subsection{Performance evaluation criteria}
We use Mean Absolute Error (MAE), and R-squared ($R^2$) metrics to evaluate those models' performance. MAE is defined as the average of the absolute difference between the target values and the values predicted by the model. 

\begin{equation}
M A E=\frac{1}{n} \sum_{i=1}^{n}\left|\hat{y}_{i}-y_{i}\right|
\end{equation}

$R^2$ is defined as the part of the variance of the dependent variable based on the independent variables of our model.

\begin{equation}
R^{2}=1-\frac{\sum_{i}^{n}\left(y_{i}-\hat{y}_{i}\right)^{2}}{\sum_{i}^{n}\left(y_{i}-\bar{y}\right)^{2}}
\end{equation}

In our work, k-Fold Cross-Validation is used to evaluate the performance of trained machine learning models on the limited dataset. Compared to random training test splitting, k-Fold Cross-Validation allows us to obtain less biased, less optimistic, and more stable estimates of model performance. The basic procedure is the dataset is shuffled randomly and split into k number of groups, then take each group in turn as the test dataset, then take the remaining groups as the training dataset. Fit the model on the training set and evaluate it on the test set. The performance of each of the trained models are averaged and reported as the overall cross validation performance. In our evaluations, we used 10-Fold Cross-Validation to ensure fair and stable results.

\section{Results}
\label{sec:others}

\subsection{Global distribution of piezoelectric materials}

\begin{figure}[!th]
  \begin{subfigure}[t]{0.99\textwidth}
    \includegraphics[width=0.9\textwidth]{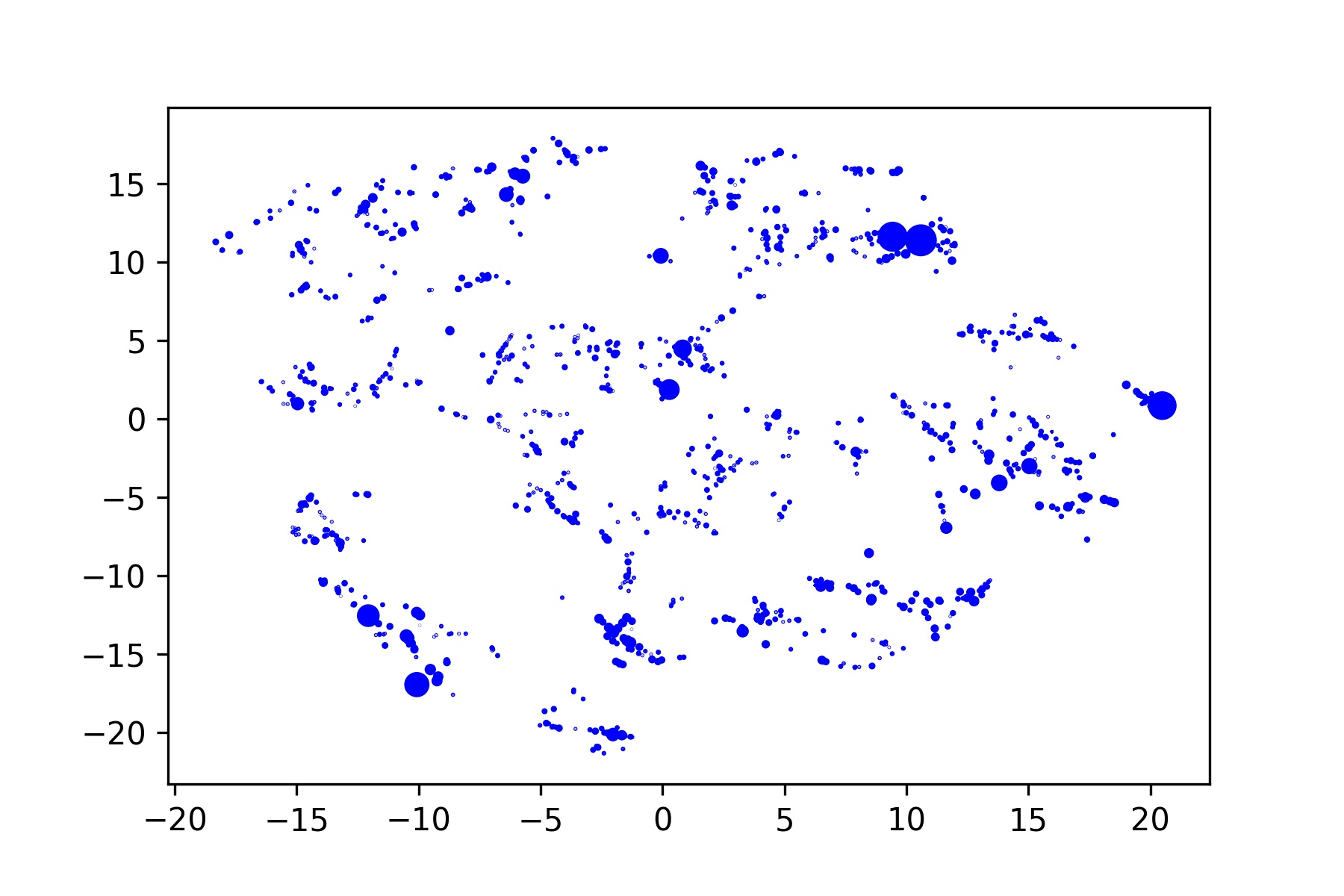}
    \caption{T-sne map of piezoelectric moduli.}
    \label{fig:map1}
  \end{subfigure}
  \hfill
  \begin{subfigure}[t]{0.99\textwidth}
    \includegraphics[width=0.9\textwidth]{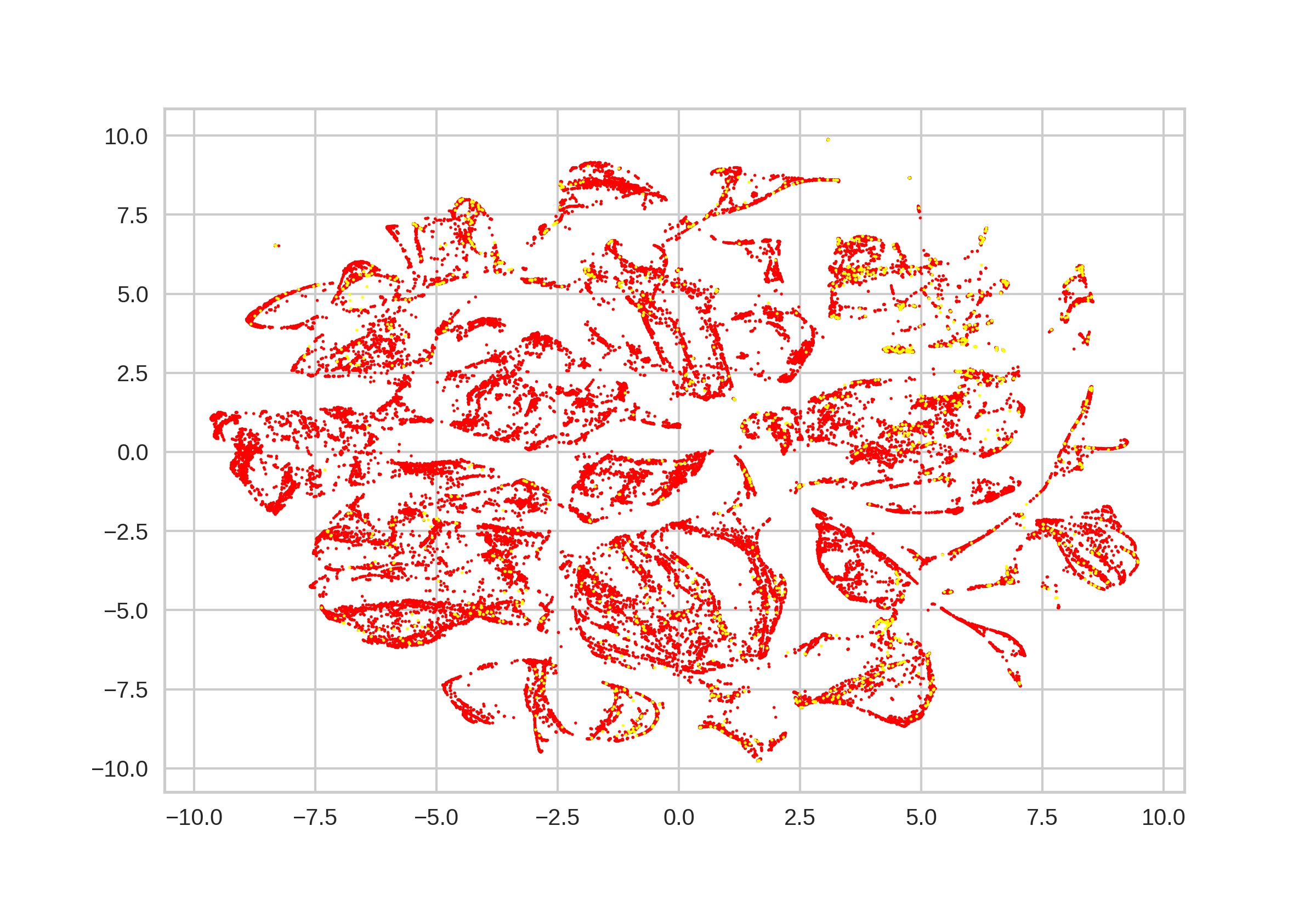}
    \caption{Distribution of 1,705 known piezoelectric materials vs other ternary materials.}
    \label{fig:map2}
  \end{subfigure}
  \caption{Distribution of piezoelectric moduli and materials in composition space. (a) the sizes of the points are proportionate to their piezoelectric moduli. There are a few exceptionally high values located in 4 clusters. (b) Ternary materials (red points) are distributed in composition clusters. Known piezoelectric materials (yellow points) are similarly grouped in clusters with two clusters with a significant number of known materials. }
\end{figure}

\FloatBarrier
To explore how the piezoelectric materials and their piezoelectric modulus are distributed, we visualize our piezoelectric dataset with 1,705 samples all using the t-sne algorithm \cite{van2008visualizing}. This algorithm is capable of mapping high dimension data points into 2D space by preserving the neighborhood relationships. Each of the samples are first featurized by calculating their magpie features, which are then fed to the t-sne algorithm that maps them into 2D space. Two distribution maps are generated using this procedure. We only include the 1,705 materials samples with piezoelectric modulus values. The map is shown in Figure \ref{fig:map1} with blue colors and their size indicates high-piezoelectic effect. We find that only a small number of materials have higher piezoelectric properties.

To further show how the know piezoelectric materials are distributed among all ternary crystal materials, we include all ternary materials samples (with the 1,705 piezoelectric materials included). The distribution map is shown in Figure \ref{fig:map2}. Red dots are the 59,211 ternary materials samples and blue dots are the 1,705 piezoelectric materials. It can be observed that the overlap area between red and blue dots is very small indicating that piezoelectric materials are different from general materials distribution.


\subsection{Performance of machine learning models}
The performances of our Random forest models and SVM models with different types of features (magpie features, oxidation states features, structural features, feature transformations, energy and magnetism features, and elastic modulus) are reported in Table \ref{tab:performance_table}.

\begin{table}[htb!]
\centering
\caption{Machine learning performances on piezoelectric constant prediction (10-fold cross-validation).}
\label{tab:performance_table}
\begin{tabular}{|l|c|c|c|c|}
\hline
 & \multicolumn{1}{l|}{\textbf{RF ($R^2$)}} & \multicolumn{1}{l|}{\textbf{RF (MAE)}} & \multicolumn{1}{l|}{\textbf{SVM ($R^2$)}} & \multicolumn{1}{l|}{\textbf{SVM (MAE)}} \\ \hline
Magpie Features & -0.509 & 1.17 & 0.043 & 0.841 \\ \hline
Magpie Features + Oxidation States Features & -0.480 & 1.18 & 0.047 & 0.840 \\ \hline
Magpie + Oxidation States + Structural Features & -0.314 & 1.026 & 0.094 & 0.764 \\ \hline
\shortstack{Magpie + Oxidation States + Structural Features\\+ Feature Transformations}
& -0.334 & 1.017 & 0.110 & 0.750 \\ \hline
\shortstack{Magpie + Oxidation States + Structural Features\\ + Energy and Magnetism Features} & -0.385 & 1.061 & 0.114 & 0.748 \\ \hline

\shortstack{Magpie + Oxidation States + Structural Features\\ + Energy and Magnetism Features\\+ Elastic Modulus Features} & -0.343 & 0.953 & 0.127 & 0.646 \\ \hline

\end{tabular}
\end{table}

Initially, the random forest model is only based on the magpie features, with its MAE and $R^2$ scores being 1.17 $C/m^2$ and -0.509. After adding the oxidation state features, structural features, energy and magnetism features, and elastic features, the MAE decreases 18.5\% to 0.953  and the $R^2$ increases 32.6\% to -0.343. 

The SVM's $R^2$ performance score starts at 0.043 with magpie features and then increases 117\% (more than doubled) with the addition of oxidation states and structural features. This shows the importance of structural information for piezoelectric modulus prediction. Unlike the random forest model, the SVM model performs better with the addition of energy and magnetism features (increased 22\%) showing that SVM is more stable than the random forest models. The MAEs for both random forest and SVM show moderate change with the addition of the diverse features showing that MAE has more potential to improve  (e.g. with undiscovered features). As elastic moduli are important quantities that measure the resistance to being deformed elastically, here we also add the bulk and shear moduli to train the random forest model and the SVM model. The MAE and $R^2$ of the random forest model improve 10.17\% and 10.9\% respectively while the SVM model’s MAE and $R^2$ improve 13.63\% and 11.4\% accordingly.

\begin{figure}[htb!]
\centering

\begin{subfigure}[b]{0.75\textwidth}
   \includegraphics[width=1\linewidth]{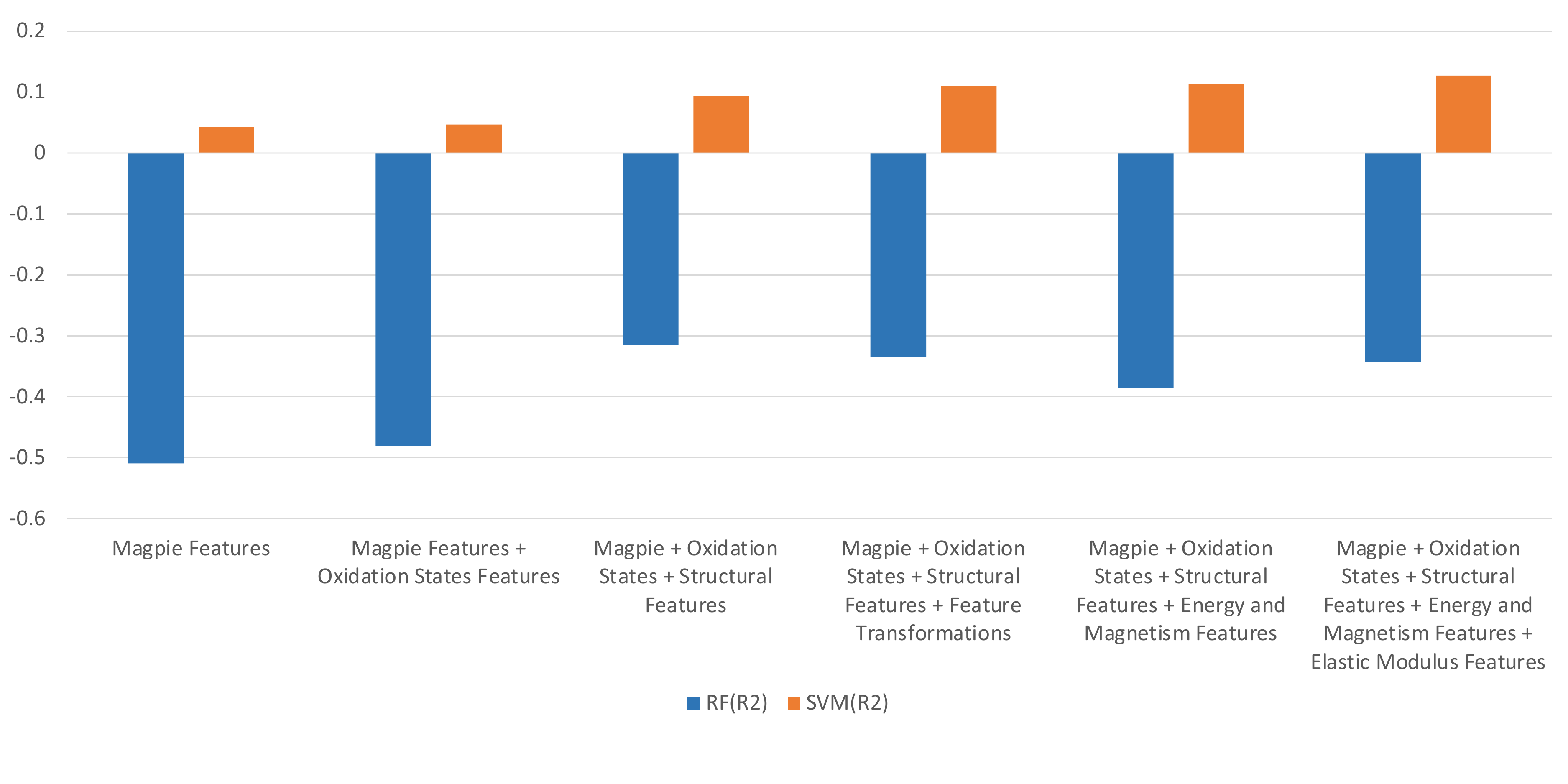}
   \caption{$R^2$ score}
   \label{fig:R2_performance}
\end{subfigure}

\begin{subfigure}[b]{0.75\textwidth}
   \includegraphics[width=1\linewidth]{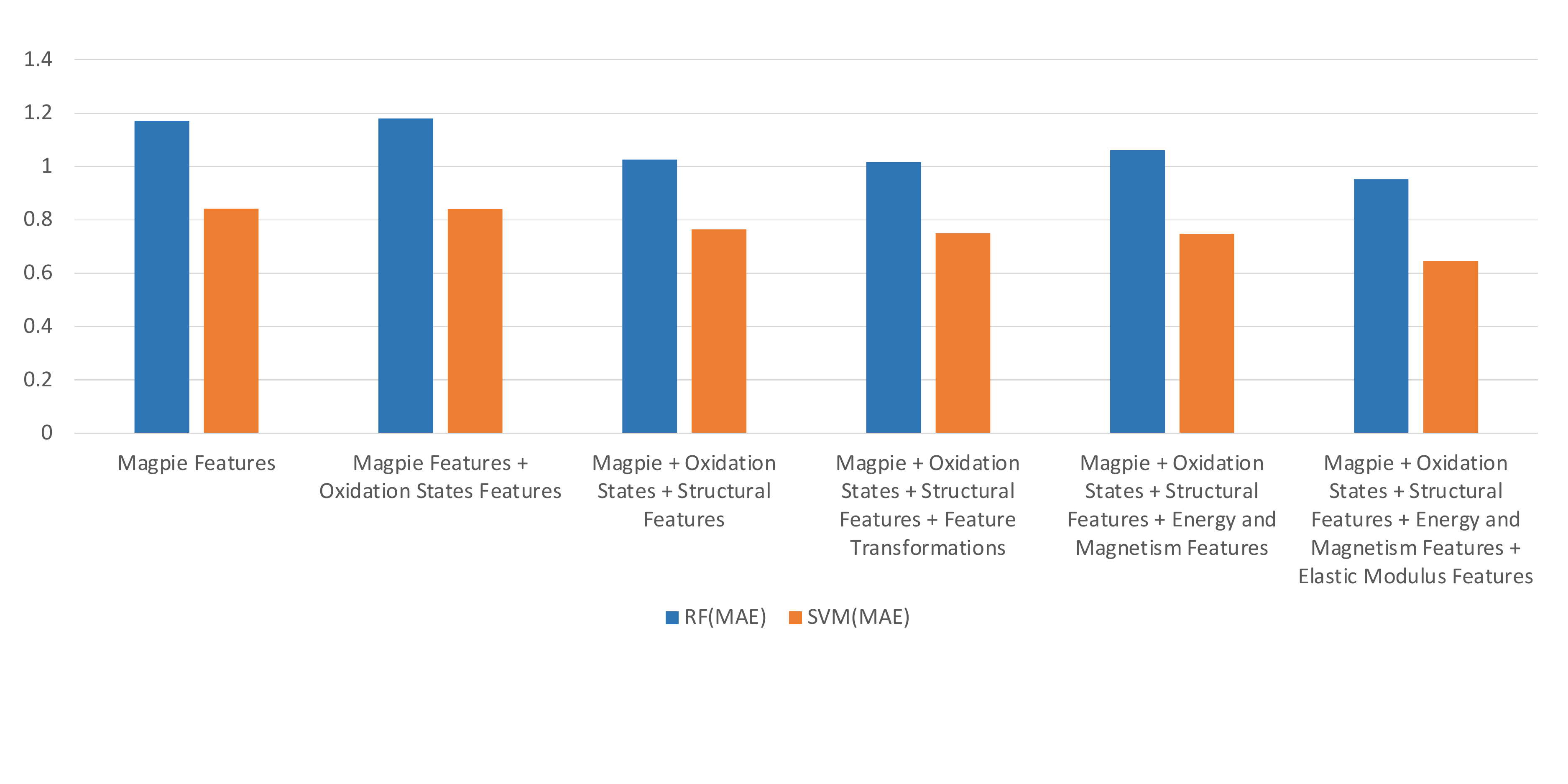}
   \caption{MAE error}
   \label{fig:MAE_error} 
\end{subfigure}

\caption{Performance comparison of random forest and SVM in terms of $R^2$ scores and MAE errors.}
\label{fig:histograms} 
\end{figure}

For better comparison, we draw two bar charts in Figure \ref{fig:histograms} to show the performance changes with different combinations of features. Figure \ref{fig:R2_performance} shows the $R^2$ performance where the blue bars represent the $R^2$ scores of the random forest models and the red bars represent those of the SVM models with different feature sets. From the figure, we find that the SVM's $R^2$ scores remain positive throughout the six different feature sets whereas the random forest's $R^2$ scores remain negative, indicating that SVM models perform better than random forests in terms of the $R^2$ performance measure. As more features are added throughout the feature sets, SVM consistently increases its accuracy. Random forest, instead, shows different patterns. Starting with the magpie features, adding oxidation states and structural features increases the accuracy whereas adding feature transformations and energy and magnetism features lowers the accuracy score. 
In Figure \ref{fig:MAE_error}, the blue bars represent the MAEs of the random forest model and the red bars represent the MAEs of the SVM models for piezoelectric prediction. In the figure, we notice that the MAE for the random forest model (blue bars) is significantly greater than the MAE for SVM models (red bars) indicating random forest performs worse than SVM. Another pattern we find is that as more features are added, the error for both random forest and SVM decreases.

\FloatBarrier

\begin{figure}[htb!]
\centering
  \begin{subfigure}[b]{0.49\textwidth}
      \centering \includegraphics[width=0.99\linewidth]{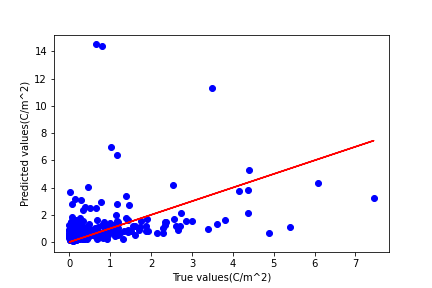}
  \caption{RF prediction with with five types of features }
  \label{fig:piezo_rf_scatter}
  \end{subfigure}
\begin{subfigure}[b]{0.49\textwidth}
      \centering \includegraphics[width=0.99\linewidth]{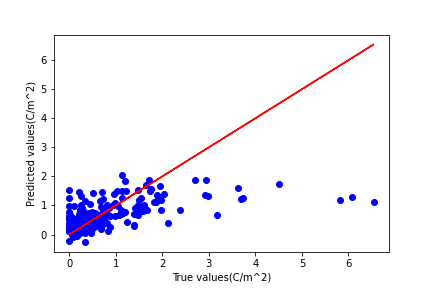}
  \caption{SVM prediction with five types of features}
  \label{fig:piezo_svm_scatter}
  \end{subfigure}
  
  \caption{Scatter plots of predicted piezoelectric coefficients by random forest model and SVM model }
  \label{fig:scatter}
\end{figure}

To compare the predicted piezoelectric coefficients with the true piezoelectric coefficients, we draw two scatter plots in Figure \ref{fig:scatter}, in which the x-axis indicates the true value, and the y-axis shows the predicted value. In addition, the red color line is the regression line. 
The results of the random forest model with five types of features are shown in Figure \ref{fig:piezo_rf_scatter} and results of the SVM model are shown in Figure \ref{fig:piezo_svm_scatter}. The more blue dots close to the red line, the better performance the model achieves. For RF model, more blue dots are far away from the red regression line than SVM, which indicates that the performance of SVM is better. Besides, there are several dots with higher predicted piezoelectric coefficients (around 14) while their true values are small (around 1). For the SVM model, the overall fitting within 2 $C/m^2$ is good, which is consistent with the fact that the range of most piezoelectric coefficients are between 0 to 2 $C/m^2$ (refer to Figure \ref{fig:histograms}). 

\subsection{Performance of graph neural networks}

Table \ref{tab:gnn_performance} shows the five fold cross validation results of piezoelectric modulus prediction by graph neural networks. The best graph neural network model is CGCNN with a MAE of 0.97439 C/$m^2$ and the worst model is SchNet with a MAE of 1.34294 C/$m^2$. Compared to random forest and SVM in Table \ref{tab:performance_table}, every graph neural network under performed the SVM model. SchNet, CGCNN, MPNN, and MEGNET all performed slightly better than random forest while GATGNN performed slightly worse.

\begin{table}[]
\centering
\caption{Performance of graph neural network for piezoelectric coefficient prediction. The MAE errors are much larger than those of SVM models.}
\label{tab:gnn_performance}
\begin{tabular}{|l|l|l|l|l|l|}
\hline
Model & SchNet & CGCNN & MPNN & MEGNET & GATGNN \\ \hline
MAE (C/$m^2$)& \multicolumn{1}{r|}{1.34294} & \multicolumn{1}{r|}{0.97439} & \multicolumn{1}{r|}{1.04362} & \multicolumn{1}{r|}{0.98129} & \multicolumn{1}{r|}{1.09634} \\ \hline
\end{tabular}
\end{table}



\subsection{Feature analysis}

To obtain physical insights from our machine learning models and the dataset, we analyze the distribution of the piezoelectric materials in terms of the crystal systems and also conducted the feature importance analysis to identify physical factors that affect materials' piezoelectric moduli.

\begin{figure}[htb!]
  \centering
      \includegraphics[width=0.65\textwidth]{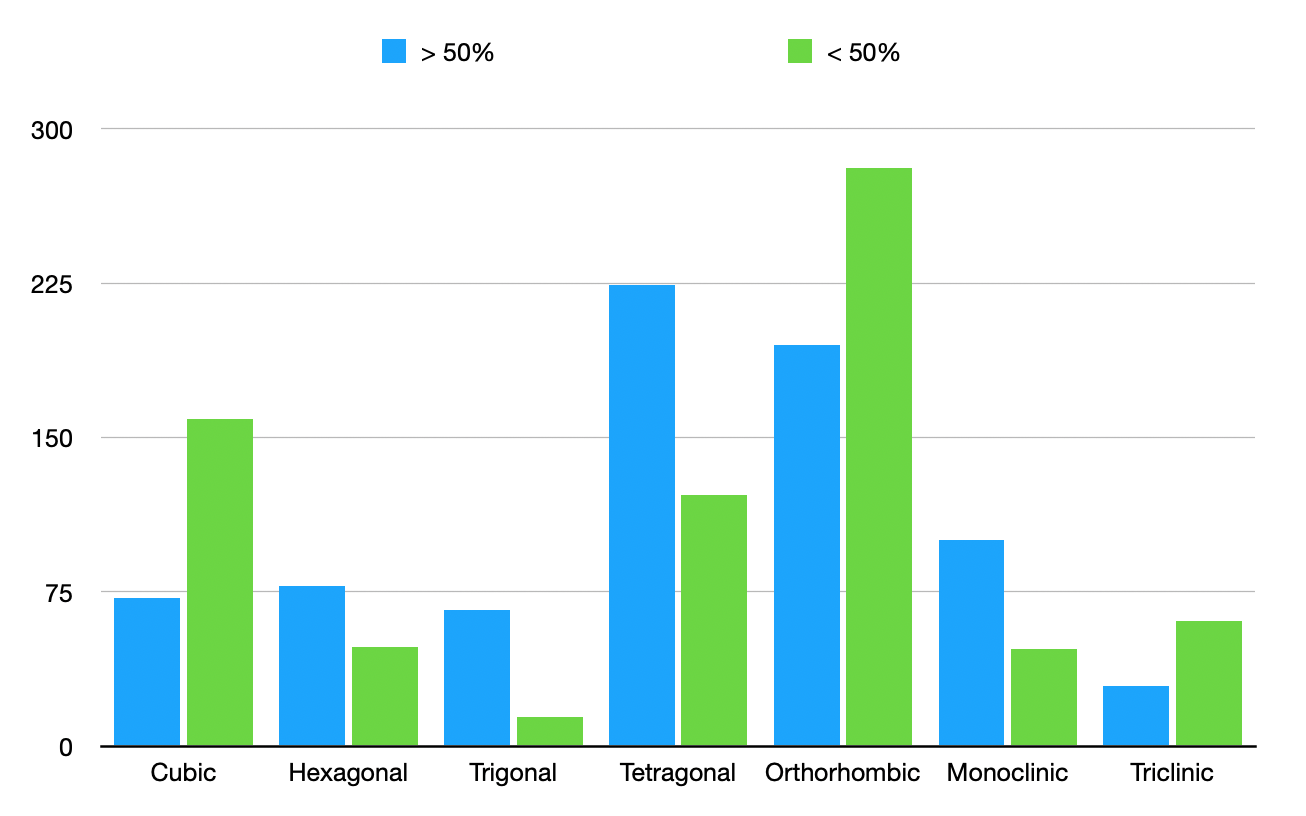}
  \caption{Distribution of the number of materials in different crystal systems with high (>50\% percentile) and low( <50\% percentile) piezoelectric modulus. It is found that materials with high piezoelectric modulus tend to belong to crystal systems with intermediate symmetry (tetragonal, orthorhombic, monoclinic).}

  \label{fig:piezo_distribution}
\end{figure}

From Figure \ref{fig:piezo_distribution} we can see that the distribution of known piezoelectric materials is highly unbalanced among the seven different crystal systems. The system with the highest number of known piezoelectric materials (frequency) is orthorhombic, followed by tetragonal and cubic. The remaining four crystal systems have much fewer known piezoelectric materials. It seems that the structure property has profound influence on the piezoelectric effect. We suspect that the reason that there are a smaller number of monoclinic and triclinic piezoelectric materials exist is the low symmetry of the crystal structure. We also observed the high symmetry crystal systems tend to also have a smaller number of known piezoelectric materials. We find that the cubic crystal system has 231 piezoelectric materials despite its high symmetry. This may be due to the fact that cubic materials are from the most well-studied material family. Based on this analysis we believe structure features are highly important in building high performance models for piezoelectric modulus.

\begin{figure}[htb!]
  \centering
      \includegraphics[width=0.95\textwidth]{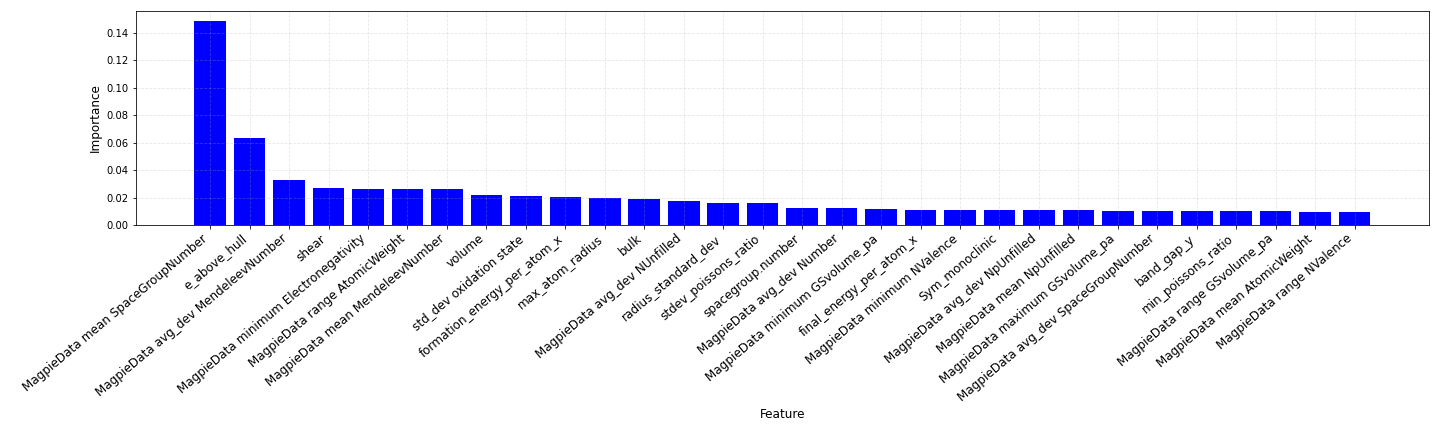}
  \caption{Ranking of top 30 features in the RF model for piezoelectric modulus prediction. It brings physical insights by showing the key physical factors that affect the piezoelectric effect.}

  \label{fig:importance}
\end{figure}

To further understand how different types of features contribute to the final prediction performance, we trained the random forest model which provides a built-in ranking of the feature importance. The top 30 features are shown in Figure \ref{fig:importance}. 

The most important feature is the mean SpaceGroupNumber followed by energy above hull, average MendeleevNumber, shear and minimum Electronegativity. Since element space group number represents the symmetry trends of crystal materials, it can explain why this feature is important. As for shear  

In terms of Poissons' ratio, it reflects the degree of volume expansion perpendicular to the applied force which is closely related to the piezoelectric effect. It's interesting that our random forest mode can identify Poisson's ratio as one of the top features of the piezoelectric modulus prediction model, which is also confirmed in experimental study \cite{ogi2004elastic}.

Another two important features are related to Nf valence and Np valence, which is related to how easy the atoms lose electrons. This can explain why this set of valence related features are ranked high in this prediction model. We also found the space group number of the crystal material which is ranked as the 17th most important feature along with symmetry orthorhombic ranked 26th. This is consistent with our analysis in Figure \ref{fig:piezo_distribution} showing that crystal structure plays a crucial role in the distribution of piezoelectric modulus. We also found that total magnetization is ranked 25th which explains why we obtain the better results when we add these features as shown in Figure \ref{fig:R2_performance}. Next, we also found that random forest ranked the density features 28th and 29th, consistent with the fact that high density materials are not easy to deform and thus tend to have lower piezoelectric modulus. Lastly, we found that Nd unfilled feature is ranked 30th. This feature represents the capability for the atom to acquire electrons so that it may affect the piezoelectric effect. 

\FloatBarrier

\subsection{Predicted potential piezoelectric materials}

To discover new potential piezoelectric materials, we applied our trained SVM model with five kinds of features to predict the piezoelectric properties of 12,680 materials from the Materials Project database which have elastic modulus values (bulk and shear). We sort the predicted piezoelectric coefficients and find 12,650 materials' predicted values are positive with 1,498 of them are larger than 1 $C/m^2$ and 12 materials with predicted piezoelectric coefficients greater than 2 $C/m^2$. We listed the top 20 predicted piezoelectric materials in Table \ref{tab:top20_table}. We also summarize a total of 150 materials with high predicted piezoelectric coefficients in supplementary file Table S1, Table S2 and Table S3, which are classified according to the number of elements in the materials.

\begin{table}[htb!]
\centering
\caption{Top 20 predicted potential piezoelectric materials (unit: $C/m^2$)}
\label{tab:top20_table}

\begin{tabular}{|c|c|c|c|c|c|}
\hline
Material ID & Formula & Predicted Value  &
Material ID & Formula & Predicted Value   \\ \hline
mp-578601                          & NaNbO2                               & 2.448                          & mp-756683                         & HfBiO4                               & 2.015                          \\ \hline
mp-10426                           & Nb2O5                                & 2.341                          & mp-7017                           & NaNbN2                               & 2.007                          \\ \hline
mp-760401                          & Nb3O7F                               & 2.184                          & mp-549490                         & KNb4O5F                              & 1.978                          \\ \hline
mp-754698                          & NbO2                                 & 2.153                          & mp-552588                         & LiNbO3                               & 1.966                          \\ \hline
mp-753459                          & Nb3O7F                               & 2.140                          & mp-7240                           & NaRuO2                               & 1.958                          \\ \hline
mp-1595                            & Nb2O5                                & 2.114                          & mp-754375                         & NaTi2O3                              & 1.949                          \\ \hline
mp-557680                          & NbAgO3                               & 2.085                          & mp-505517                         & BaNb4O6                              & 1.938                          \\ \hline
mp-755690                          & NbO2                                 & 2.059                          & mp-644497                         & BaTiO3                               & 1.935                          \\ \hline
mp-753380                          & La(BiO2)2                            & 2.058                          & mp-28254                          & LiRuO2                               & 1.932                          \\ \hline
mp-2533                            & NbO2                                 & 2.024                          & mp-1029267                        & CaZrN2                               & 1.928                          \\ \hline
\end{tabular}
\end{table}

\section{Conclusions}

In this study, we developed and applied two machine learning algorithms (Random forest and support vector machines) and five graph neural network models for predicting the piezoelectric modulus using a variety of composition and structural features. Extensive evaluations have been done over the dataset composed of 1,705 samples downloaded from the Materials Project data repository. Our experiment results show that the composition only descriptors/features alone do not help to build a good piezoelectric modulus prediction model, which proves that the piezoelectric
effect is strongly affected by their crystal lattice structures as well as other electronic and magnetic properties, and elastic moduli. By adding the structural features, magnetic features and elastic features, we have been able to increase the regression $R^2$ score of the SVM model from 0.043 to 0.127. Our study also shows that the Random forest models in general perform much worse than the SVM models. We also explored the popular graph neural network models for the piezoelectric modulus prediction, which shows that it is much more challenging and all these models have much lower performance compared to those of other property prediction \cite{fung2021benchmarking}, which may be due to the limited dataset size and the sophisticated relationships between crystal structures and the piezoelectric effect. However, our study does show that the graph neural network models can achieve performance compared to RF and SVM in terms of the MAE errors. Compared to other properties such as formation energy, shear/bulk moduli, and band gaps, we find that piezoelectric modulus is notoriously much more challenging to predict accurately. Much more detailed feature engineering or deep neural network based representation learning are needed to further improve the prediction performance of piezoelectric modulus to enable large-scale screening and design of novel piezoelectric materials and uncover the complex relationships of the crystal structure and the piezoelectric effect. More importantly, we utilize the trained SVM model to predict the piezoelectric coefficients for 12,680 materials from the materials project database, and find 1,498 materials whose predicted piezoelectric coefficients are lager than 1 $C/m^2$. We report the top 20 materials with all related information (material ids in material project dataset, formulas and the predicted piezoelectric coefficient values), which could inspire experimental material scientists to verify some of these new piezoelectric material candidates.


\bibliographystyle{unsrt}  
\bibliography{references}  






\end{document}